# Deep Autoencoder for Recommender Systems: Parameter Influence Analysis

**Dai Hoang Tran**
Department of Computing
Macquarie University
Sydney, Australia
Email: dai-hoang.tran@hdr.mq.edu.au

**Zawar Hussain**
Department of Computing Macquarie University
Sydney, Australia
Email: zawar.hussain@hdr.mq.edu.au

**Wei Emma Zhang**
Department of Computing Macquarie University
Sydney, Australia
Email: w.zhang@mq.edu.au

**Nguyen Lu Dang Khoa**
CSIRO Data61
Sydney, Australia
Email: khoa.nguyen@data61.csiro.au

**Nguyen H. Tran**
School of Information Technologies
The University of Sydney
Sydney, Australia
Email: nguyen.tran@sydney.edu.au

**Quan Z. Sheng**
Department of Computing Macquarie University
Sydney, Australia
Email: michael.sheng@mq.edu.au

## Abstract

Recommender systems have recently attracted many researchers in the deep learning community. The state-of-the-art deep neural network models used in recommender systems are typically multilayer perceptron and deep Autoencoder (DAE), among which DAE usually shows better performance due to its superior capability to reconstruct the inputs. However, we found existing DAE recommendation systems that have similar implementations on similar datasets result in vastly different parameter settings. In this work, we have built a flexible DAE model, named FlexEncoder that uses configurable parameters and unique features to analyse the parameter influences on the prediction accuracy of recommender systems. This will help us identify the best-performance parameters given a dataset. Extensive evaluation on the MovieLens datasets are conducted, which drives our conclusions on the influences of DAE parameters. Specifically, we find that DAE parameters strongly affect the prediction accuracy of the recommender systems, and the effect is transferable to similar datasets in a larger size. We open our code to public which could benefit both new users for DAE-they can quickly understand how DAE works for recommendation systems, and experienced DAE users- it easier for them to tune the parameters on different datasets.

# 1 Introduction

A recommender system is a system which recommends certain items to its users and those recommended items yield a better user response than the non-recommended items. Nowadays, recommender systems are playing an important role in the modern technology services. They help boost business and facilitate decision making for many companies in different industries such as Amazon book recommendation[1], Spotify music recommendation[2], Netflix movie recommendation[3] and Google Play Store mobile application recommendation[4]. Current technology services often provide their users the most relevant content to enhance engagement and reduce users' effort in finding the content of their interests. There are two major approaches for traditional recommender systems. The first one is Collaborative Filtering (CF) (Bell and Koren 2007), in which each user is recommended with items based on other users with similar preferences. The second approach is using content-based recommendation (Liu et al. 2010), in which each user is recommended with items that are similar to what he/she liked previously. Some techniques use a hybrid approach by combining both collaborative filtering and content-based methods (Burke 2002). All of these approaches perform well in many applications. However, they usually face certain limitations and challenges due to the increasing demand of high quality personalization and recommendation (Adomavicius and Tuzhilin 2005).

Recent developments have seen the rising of adopting the neural network model for recommender systems. There is a big focus on using Autoencoder to learn the sparse matrix of user/item ratings and then perform rating prediction (Hinton and Salakhutdinov 2006). A large body of research works has been done on Autoencoder architecture, which has driven this field beyond a simple Autoencoder network. Many different techniques have been proposed such as *denoising* architecture or *dropout* to improve the system effectiveness (Kuchaiev and Ginsburg 2017) and combination of neural network with collaborative filtering methods (Strub et al. 2016) to improve prediction. Besides the differences in theory, all these techniques have quite similar implementations of the DAE network but there is no agreement on optimal parameters used to train the model efficiently. Each work has used different set of hyper-parameters and specific settings but the implementations are similar.

This has motivated us to build a flexible deep neural network model, where we can embed our features and techniques from other works, into one comprehensive DAE model, called FlexEncoder. FlexEncoder contains more than 15 configurable parameters for tuning, and can significantly affect the rating predictions. Our goal is not only to build a feature-rich Autoencoder model for recommender systems, but also to find out what kind of settings and what set of parameters will provide a better recommendation. By conducting an extensive evaluation on our FlexEncoder model, we want to answer an important question: How do parameters of DAE network influence the rating prediction?

Our main contributions in this work are as follows:

- We provide a fully open-source and feature-rich DAE network for recommender systems[5]. This will help new researchers to quickly understand the Autoencoder -implemented recommender systems and test per their request.

- We implement unique features into our model, such as *prediction rounding* or *dense re-feeding rounding* (detailed in Section 3.1). This will help to enrich the set of features and ideas in the research area of Autoencoder recommender systems.

- We perform an extensive evaluation of our model with various parameter's combinations, which help us to understand the effect of Autoencoder parameters on influence the rating prediction.

# 2 Background

Recommender system is a well-known research field for the last few decades (Hill et al. 1995). It has gained huge attraction from industry in the modern age due to the problem of information overload. In recent developments, researchers are applying deep neural network to make recommender systems more robust and to overcome the challenges such as cold-start problem and sparse rating matrix (Adomavicius and Tuzhilin 2005). In this section, we will introduce recommender systems with an industry user case, introduce its fundamental models and challenges.

---

[1] https://www.amazon.com

[2] https://www.spotify.com

[3] https://www.netflix.com

[4] https://play.google.com

[5] *https://github.com/heroddaji/SurpriseDeep*

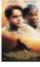

*Figure 1: A Movie Recommender System*

## 2.1 A User Case of Recommender Systems

To understand what a recommender system is, we take the e-commerce service Amazon[6] as an example. Amazon lists and sells millions of items such as books and accessories to customers around the globe. Due to its massive size of catalogues and large user base, Amazon cannot show all of its products to a customer via the website. Instead, they want to tailor their services to each customer based on his/her purchasing history and pattern while browsing the website. As a result, each customer who visits Amazon, will see such items that are "recommended" to fit that user's taste, hence increasing the chance of that user to buy more items on Amazon. Therefore, the fundamental idea of a recommender system is to leverage difference sources of data to infer or predict customer interests.

For the terminologies, the products to be recommended are called *items*, and the entity to receive the recommendation is the *user*. The choice of recommended items is generally based on *rating prediction value*, in which *rating* is the user feedback on certain items. Figure 1 shows an example of a classical movie recommender system.

In Figure 1, we have five *users*, and each of them provides ratings for different movies, which are the *items* in this case. A rating value is the number of stars from total of 5 stars, thus the *rating* values are in range of 1 to 5. The objective of this system is to predict the missing ratings of each user, as you can see with red question marks. Additionally, the table containing the rating data is called the *rating matrix*. One particular interested property of a recommendation rating matrix is that it is very sparse. That means there are lots of missing entries, and dealing with sparse matrix is one of the fundamental challenges in recommender systems. The example we illustrate here is a very simple recommender system. In the real world, there are large variety of products and recommender systems for different purposes such as music recommendation (Patrick et al. 2017) and news personalization (Domingos and Richardson 2001).

## 2.2 Fundamental Models and Challenges of Recommender Systems

Most recommender systems are designed to work with two types of data, which are (i) the information of the users and items such as user profiles, user locations, item description or item keywords, and (ii) the interactions between item-user such as rating values. The models that use the first type of data are regarded as *content-based recommender* models, whereas models what use the second type of data are considered as *collaborative filtering* models. There are some recommender systems which leverage both data types and techniques to make a *hybrid* system (Adomavicius and Tuzhilin 2005).

- **Content-based recommendation**: The user will be recommended items similar to the ones that the user has preferred in the past, for instance: based on their own previous purchase history. For example, a user has bought several history books in the past. Thus, in her next visit to the online bookstore, she is recommended with history books. The system uses "description" data in those books to infer her interests.
- **Collaborative filtering recommendation**: The user will be recommended items based on other users who have similar tastes and preferences. For example, user *A* rated certain action genre movies highly and the system wants to infer whether user *A* will have any interest in romantic genre movies. To be able to do that, the system will find similar users who have given high ratings for action genre movies and have also given ratings to romantic genre movies. Then by combining those ratings from other similar users, the system can infer user *A's* interest in romantic genre. The idea is that users who share similar interests may behave in a similar manner.

---

[6] https://www.amazon.com

Regardless of which recommendation models being used, there are certain challenges that a good recommender system would encounter. We list them as the following (Adomavicius and Tuzhilin 2005):

- **New user problem**: For a user to receive accurate recommendation, he/she needs to rate a sufficient amount of items before the recommender system can understand his/her preferences. As a result, new users who have few ratings, will get low accuracy recommendation.
- **New item problem**: The problem stems from the fact that collaborative methods work only on other users' preferences and previously rated items to make recommendation. As a result, new item which has not been rated yet, will not get recommended until some users discover and rate it.
- **Memorization and generalization**: Content-based recommender systems usually suffer from high *memorization* rate problem, i.e., users are recommended with only those items that are similar to their previously rated items, thus lacking the diversity. However, diversity or *generalization* of the recommended items is very important, which allows the user to explore more options and further improving the user's engagement. Ideally, the user should receive a range of diverse items and not only from homogeneous categories.
- **Sparsity**: In recommender systems, it often occurs that the number of rated items is very small as compared to the number of items that have not been rated. Therefore, making accurate rating based on small samples is a very challenging task.

## 3 FlexEncoder

In this section, we review the existing neural network models for recommendation system which motivate us for developing this work (Section 3.1). Then we introduce our flexible implementation of DAE (Section 3.2).

### 3.1 Motivation

To overcome challenges mentioned in Section 2.2, researchers are now exploiting neural network models for recommender systems. Particularly, the DAE model is the choice of several researchers because of its superior capacity in learning the reconstruction of rating matrix. Autorec (Sedhain et al. 2015) is the pioneer work in this area. In this work, the authors developed I-Autorec and U-Autorec for making prediction from user-based or item-based ratings, respectively, using one hidden bottle neck layer. Kuchaiev and Ginsburg (2017) applied dropout technique while Strub and Mary (2015) introduced denoising feature into their model. Suzuki and Ozaki (2017) applied a hybrid model using both Autoencoder model and collaborative filtering method to calculate the hidden similarity for more serendipitous recommendation.

These researches provide promising models but what we have found so far is that besides the similarity of using Autoencoder model, all of these researchers are using very different parameter settings on similar datasets such as MovieLens or Netflix. Autorec used a fix set of parameters and one hidden layer. Kuchaiev and Ginsburg (2017) did experiment with different parameters, but they limited to only activation functions, optimizer and hidden layers. Similarly, Strub and Mary (2015) also used one fix set of parameters with TANH activation function and SGD optimizer. We believe that understanding and tuning these parameters correctly, will have a huge impact on the correctness and robustness of the DAE model. This is the motivation behind this work to develop our own DAE model for recommender systems, called FlexEncoder.

### 3.2 Architecture

Our FlexEncoder model is a deep Autoencoder. An Autoencoder is a neural network that has two functions during the training process, namely the encode function $encode(x): R^n \to R^d$ and the decode function $decode(z): R^d \to R^n$. The main objective of Autoencoder is to retrieve the data representation of $d$ dimension so that the error between $x$ and the reconstruct function given in Equation 1, is minimized (Hinton and Salakhutdinov 2006). This is the basic model for DAE network. There are other variants such as denoising Autoencoder (Vincent et al. 2008) in which additional noises are added to the inputs to make hidden layers be able to discover more latent features and provide more robust results. Our model follows denoising Autoencoder architecture:

$$f(x) = decode\big(encode(x)\big) \qquad (1)$$

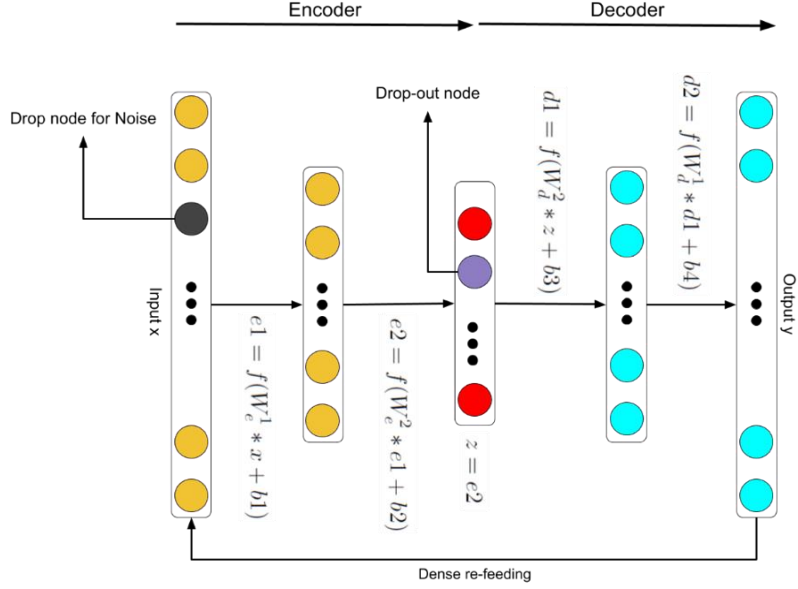

*Figure 2: The FlexEncoder Model*

Figure 2 shows one example of our model with three hidden layers. Notice that the input layer can drop certain nodes for noise corruption. The bottleneck layer $z$ also has *dropout* technique applied. Aside from that, FlexEncoder model is similar to feed-forward neural network with fully connected layers and each layer makes the computation of value $c$ as given by Equation 2:

$$c = f(W * x + b) \qquad (2)$$

Here $f$ is the activation function, $W$ is the weight, $x$ is the input value and $b$ is the bias. The decoder part is the reversed architecture of the encoder part and it comes after the bottleneck layer in the middle. There are certain manipulation techniques that can be applied to the decoder part, such as constrained decoder. If constrained decoder is used, the decoder weights $W_d^l$ will be equal to the transpose of encoder weights $W_e^l$, and this restricts the weights freedom of the decoder.

## 3.3 Features

We implement different features originated from various works such as *denoising* feature from (Strub and Mary 2015), *dropout* from (Srivastava et al. 2014), *dense re-feeding* feature from (Kuchaiev and Ginsburg 2017) and autoencoder rating for user-based and item-based in (Sedhain et al. 2015).

Additionally, we add several experimental features to our FlexEncoder model such as:

- *Prediction rounding*: After training our FlexEncoder, we run it on the test set where we mask out certain ratings to calculate the prediction accuracy. The FlexEncoder can predict rating value ranging from -2 to 6 or 7, whereas the actual valid rating from MovieLens dataset is in the range of [1, 1.5, 2, … 4.5, 5]. This out-of-valid-range predicted ratings result is useful in many scenarios, such as when we want to get the top $k$-items with highest ratings for a target user. However, with *prediction rounding* feature enabled, any predicted ratings from our model will be rounded off to the nearest suitable rating range. For example, if the predicted rating is 2.34, it will be rounded off to 2.5 and if the predicted rating is 2.15, it will be rounded off to 2. Doing so will result in a higher number of items with highest rating (rating value 5). This technique can improve the diversity of the recommendation items, but it can affect the prediction accuracy. We will discuss this effect in the evaluation section.

- *Dense re-feeding rounding*: When this feature is enabled, d*ense re-feeding* technique is activated (Kuchaiev and Ginsburg 2017). Like the *prediction rounding*, during the *re-feeding* phase, we round off all the output values in feeding layer to the range of 1 to 5. However, our analysis shows that enabling d*ense re-feeding rounding* feature does not improve the prediction accuracy.

- *Mean normalization*: This is a process of averaging each user's ratings around value *zero*, which effectively removes the user biasness, since some users may be strict in rating than others. Each user's or movie's mean is calculated by Equation 3:

$$\mu_i = \frac{1}{n}\sum_{x=1}^{n} x_i \qquad (3)$$

where $x_i$ is the user or the movie rating. The new *normalized rating* is $\overline{r_i} = r_i - \mu_i$. Same idea of *mean normalization* is also used by (Strub et al. 2016) but we extended it by comparing both evaluations where we disable and enable *mean normalizatio*n to see its effect on the accuracy of our model.

## 3.4 Configurable Parameters

The name FlexEncoder means "flexible autoencoder". By reviewing the previous literature related to the autoencoder model, we found that each work used a very different set of hyper-parameters and other settings on the same dataset. Their choice of parameters is mostly heuristic based and are selected through trial and error approach, doing multiple experiments until a suitable set of parameters is found. This final set of parameters is then used across the whole process for the evaluation. We want to analyse these parameter settings, as we are curious to know that whether certain choices of parameter can reveal any pattern that provides good accuracy in rating prediction, and if there is, why that particular choice works well but not the others. To meet this goal, we implemented our FlexEncoder with extensive configurable parameters. Table 1 gives an explanation for all of the parameters in FlexEncoder. Some of them are hyper-parameters, while others are enhancements with a purpose to improve the prediction accuracy. Each time the model is trained, a certain set of parameters get chosen to evaluate its effect on the prediction accuracy.

For the categorical parameters such as activation types, we chose six activation functions, which are *SELU, RELU, RELU6, ELU, LRELU, SIGMOID, TANH* and SWISH. For the optimization techniques, we chose ADAM*, ADAGRA, RMSPROP* and the typical SGD. Hidden layers parameter value is selected randomly from the list of $2^1 to\ 2^{12}$ with a maximum of 5 hidden layers for the encoder. Thus, a full network can have up to a maximum of 10 hidden layers.

## 3.5 Loss Function

In a neural network, the loss function is a very crucial piece for any good model. During the training process, missing ratings are represented with value 0. It is because, the valid rating are from 1 to 5 so when we calculate and optimize the loss during backward propagation, we ignore all the zero ratings at input layer and the common loss method for that is called Masked Mean Square Error loss (MMSE). Techniques like (Sedhain et al. 2015) and (Kuchaiev and Ginsburg 2017) have also used the same MMSE loss. We also employ this method in our model implementation. In addition, the model loss measurement unit, Root Mean Square Error (RMSE), has a direct correlation with MMSE that is $RMSE = \sqrt{MMSE}$. MMSE is given by Equation 4:

$$MMSE = r_i * \frac{(y_i - q_i)^2}{\sum_{i=0}^{i=n} m_i} \qquad (4)$$

Here, $y_i$ is the real user rating, and $q_i$ is the reconstructed rating value from the model. To ignore the 0 value, we simply use a mask value $r_i$ such that $r_i = 0$, if $y_i = 0$ and $r_i = 1$ otherwise. For the special case when we enable *mean normalization* parameter, we cannot ignore rating value 0. In that case, it can be fixed by setting $r_i = 1$ for all $y_i$.

*Table 1. List of Parameters in FlexEncoder*

| Count | Parameters | Meaning | Example |
|---|---|---|---|
| 1 | **LR** - Learning rate | Learning rate $\alpha$ | 0.001 |
| 2 | **WeD** - Weight decay | Regularization term $\lambda$ | 0.001 |
| 3 | **HLs** - Hidden layers | Number of layers and dimension for each layer. For example [512, 256] means 2 hidden layers each has size of 512 and 256 nodes respectively. | [512, 256] |
| 4 | **DrP** - Drop probability | Dropout rate for bottle neck layer. | 0.3 |
| 5 | **NoP** - Noise probability | Dropout rate for introducing noises into the inputs. | 0.2 |
| 6 | **TBS** - Train batch size | One batch including many rows of record to feed into the model for every training iteration. This is the size of one batch. | 128 |
| 7 | **Ep** - Epochs | The number training cycles we repeat for one dataset. | 20 |
| 8 | **Opt** - Optimizer | The optimization algorithm to optimize the model, such as ADAM or SGD. | ADAM |
| 9 | **Act** - Activation | The type of non-linear activation function, such as RELU. | RELU |
| 10 | **DeF** - Dense re-feeding | Use *dense re-feeding* technique for k times, for example if parameter value is 1, use *dense re-feeding* one time for each training batch. | 1 |
| 11 | **DeFR** - Dense re-feeding Rounding | If enabled, *dense re-feeding* values get rounded to range of 1 to 5 | True |
| 12 | **DeC** - Decoder constraint | If enabled, put constraint on decoder weight as the transpose of encoder weight | False |
| 13 | **MeN** - Mean normalization | If enabled, apply *mean normalization* during the training and evaluation phases. | True |
| 14 | **PrR** - Prediction Rounding | If enabled, apply *prediction rounding*. | False |
| 15 | **PI** - Pivot indexes | If value is [0, 1], run model as user-based rating. If value is [1, 0], run model as item-based rating. | [0, 1] |
| 16 | **TMR** - Test masking rate | Percentage of test data get masked out (removed) for testing the model accuracy. | 0.5 |
| 17 | **TSR** - Test split rate | The pre-processing *test split rate*. | 0.3 |

# 4 Evaluation

The main objectives of our research is to find out what parameter combinations provide the optimal loss, or strike a balance between loss and training time, and also to understand the reasons behind that.

## 4.1 Dataset

We used the MovieLens 100K dataset for the evaluation of our FlexEncoder model. MovieLens 100K is managed by GroupLens and contains 100,000 movie ratings for 9,000 movies by 700 users. During the pre-processing phase, we mapped all of original user and movie identification (ID) values into a new incremental set of ID values. This helps to break any gap within the original ID values and make the computation on sparse rating matrix in our model more efficient. Also, we calculated the rating mean of each user/movie accordingly, which is called *mean normalization*. We have seen that not many previous researches have mentioned *mean normalization* in their autoencoder models or it was applied by using a fix value, such as Autorec (Sedhain et al. 2015). Autorec assigned missing entry a default rating score of 3. In our experiments, the use of *mean normalization* yields better prediction results. Other aspects such as the *split rate percent* of training and testing records are also among our parameters. The default split rate is 30% for test data but we varied it from 10% to 40% and applied randomization of records when splitting during our analysis. The rating score for MovieLens dataset is in the range of 1 to 5.

## 4.2 Setup

We have implemented our system in a flexible way using configuration files for dataset and model. Altogether, we have about 17 parameters that can be adjusted for training (Table 1). To change these parameters, we simply alter the value in the configuration files, then retrain and test the model again.

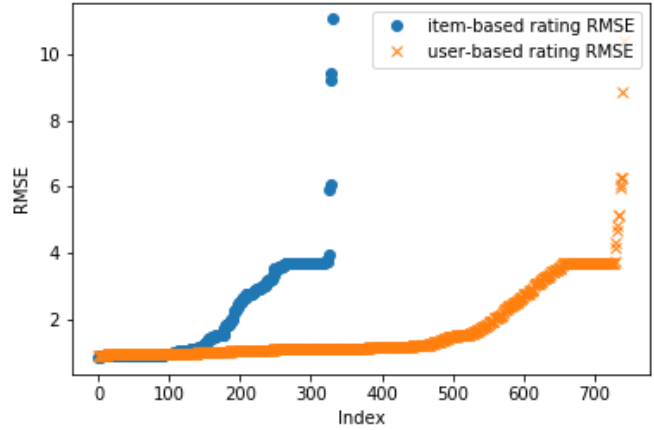

*Figure 3: Evaluation results with 700 samples for user-based and 300 samples for item-based ratings.*

When comparing our implementation to other similar works, there are many enhancements in our FlexEncoder model. Autorec (Sedhain et al. 2015) implemented the standard DAE recommender system. However, they used default value 3 to fill in the missing entry while we actually calculate the mean for each user or each movie if *mean normalization* step is enabled in configuration file. We also implemented *dense re-feeding* feature from the work of (Kuchaiev and Ginsburg 2017) and added our tweak *dense re-feeding rounding*, where we mimic the actual user input and round off them to whole or half number such as 1 or 1.5 for the re-feeding process.

All the evaluations were done with PyTorch framework [7], running on an Intel CORE i7 laptop with graphic card NVIDIA GTX 1080 and 16GB RAM.

## 4.3 Influence of Samples

From Table 1, we have 17 parameters to choose from and each of them have their own set of values. We roughly estimated that the number of possible values for all parameter combinations is more than 10 billion. Hence, it is not feasible for us to check FlexEncoder model's RMSE for all combinations. Therefore, we employed a randomized technique where each run returns a complete random set of parameters and we ran over 1000 iterations to collect the RMSE losses and analysed those results. We collected about 700 samples for user-based ratings and 300 samples for item-based ratings out of the 100K dataset with different parameter configurations. Figure 3 shows the statistics of our evaluation results. The best RMSE is 11.060 and the worst RMSE is 0.833.

*Table 2: 10 parameters sets with the lowest RMSE for MovieLens 100K rating prediction.*

| Set | HLs | WeD | DrP | NoP | Opt | Act | DeC | MeN | PrR | PI | RMSE |
|---|---|---|---|---|---|---|---|---|---|---|---|
| 1 | [64, 4] | 0 | 0.6 | 0.8 | ADAGRAD | TANH | TRUE | TRUE | FALSE | [1, 0] | 0.833 |
| 2 | [8, 32, 64, 512, 1024] | 0.001 | 0 | 0 | SGD | RELU6 | FALSE | TRUE | FALSE | [1, 0] | 0.837 |
| 3 | [4096, 4, 1024] | 0.0005 | 0.1 | 0.5 | ADAGRAD | ELU | FALSE | TRUE | FALSE | [1, 0] | 0.848 |
| 4 | [2048, 256, 32, 2] | 0.0005 | 0.5 | 0.4 | SGD | SELU | FALSE | TRUE | FALSE | [1, 0] | 0.865 |
| 5 | [16, 16, 128, 512, 1024] | 0.01 | 0.3 | 0.5 | SGD | SIGMOID | TRUE | TRUE | TRUE | [1, 0] | 0.872 |
| 6 | [2, 4, 32, 256, 2] | 0.001 | 0 | 0.5 | ADAGRAD | SELU | FALSE | TRUE | FALSE | [1, 0] | 0.877 |
| 7 | [16, 8] | 0.001 | 0.7 | 0.6 | SGD | SWISH | TRUE | TRUE | TRUE | [1, 0] | 0.879 |
| 8 | [128, 16, 256, 128] | 0.1 | 0.9 | 0.4 | ADAM | RELU6 | TRUE | TRUE | FALSE | [1, 0] | 0.884 |
| 9 | [128] | 0.01 | 0.8 | 0.4 | SGD | ELU | FALSE | TRUE | TRUE | [1, 0] | 0.885 |
| 10 | [64] | 0.1 | 0.9 | 0.5 | SGD | ELU | TRUE | TRUE | FALSE | [1, 0] | 0.886 |

---

[7] https://pytorch.org

## 4.4 Influence of Parameter Selection

The first aspect that we investigated is that what kind of parameters provide the optimal value for our main evaluation metric, the RMSE loss. Table 2 shows the top 10 combinations that provide the lowest RMSE loss. It is noticeable that all of them are item-based with *PI* value of [1, 0] and all of them have *mean normalization* feature enabled. Other parameters do not really show a pattern except the *hidden layers* which shows an interesting property: its value is either in descending or ascending order. Moreover, we found that one hidden layer can also provide a low RMSE like the multiple hidden layers. It makes us to re-evaluate our initial assumption that more hidden layers would produce better prediction accuracy for autoencoder recommender systems.

## 4.5 Overall Performance via the State-of-the-art

We also compared our model with other existing methods for the same MovieLens 100K dataset and the results are shown in Table 3. As seen from Table 3, our FlexEncoder model has the lowest RMSE as compared to other state-of-the-art recommendation techniques. The lowest RMSE score for our FlexEncoder is achieved with parameter set 1 in Table 2.

*Table 3: FlexEncoder scores lowest RMSE in comparison with other techniques. RMSE values are taken in the work of (Barbieri et al. 2017)*

| Technique | RMSE | Improvement (%) |
|---|---|---|
| **FlexEncoder (base)** | **0.833** | **0** |
| AutoRec | 0.887 | -0.064 |
| Slope One | 0.937 | -0.125 |
| Regularized SVD | 0.989 | -0.187 |
| Improved Regularized SVD | 0.954 | -0.145 |
| SVD++ | 0.903 | -0.08 |
| NMF | 0.944 | -0.133 |
| BPMF | 0.901 | -0.081 |
| RBM-CF | 0.936 | -0.123 |
| Autoencoder COFILS | 0.885 | -0.062 |
| Mean Field | 0.903 | -0.084 |

## 4.6 Analysis of Autoencoder Parameters Influence

With over 1,000 collective evaluation results from various parameter combinations, we tried to investigate, if there are any interesting patterns in the parameter combination that affects the accuracy of the model. Some general observations are already given in Section 4.3 such as *mean normalization* and item-based prediction which give quite optimal RMSE loss. Moreover, we were looking for specific patterns that are closely related to DAE network. Our first attempt was to look at the correlation between FlexEncoder parameters which is shown in Figure 4. Unfortunately, there is no clear positive or negative correlation between any pair of parameters. As a result, we have to look deeper into our experiment results, and here are some apparent patterns that we found so far:

- Our experiments with *prediction rounding* and *dense re-feeding rounding* did not yield optimal results for the prediction accuracy, when we enabled them. However, if the *prediction rounding feature* was enabled, we received more movies with 5 stars rating. This increases the diversity of recommended item content. With this observation, we plan to remove *dense-refeeding rounding* but keep *prediction rounding* experimental parameters in the future development.

- Parameter combinations that include certain *dropout* rate and *noise* rate, make the model more robust and provide good predictions. However, there is no clear evidence that a high rate of *dropout* or *noise* will make it better, since a low rate of *dropout* or *noise* result is also good in our evaluation. We conclude that *dropout* and *noise* features are necessary when applying DAE in a recommender system.

- In (Kuchaiev and Ginsburg 2017), the authors suggested that *SELU* activation function with *SGD* optimizer provided the best result for their model. However, in our case, the experimentation shows that the choice of activation functions and optimizers is uncertain for item-based rating. As shown in Table 2, different types of activation functions and optimizers can still provide a comparable low RMSE loss.

|     | LR    | WeD   | DrP   | NoP   | TBS   | TMR    | Ep     | DeF    | DeFR   | DeC    | MeN    | PrR    | TSR    | RMSE   |
|-----|-------|-------|-------|-------|-------|--------|--------|--------|--------|--------|--------|--------|--------|--------|
| LR  | 1.000 | 0.079 | 0.169 | 0.162 | 0.043 | 0.033  | -0.151 | 0.103  | 0.163  | 0.145  | -0.125 | 0.157  | -0.042 | 0.088  |
| WeD |       | 1.000 | 0.203 | 0.181 | 0.045 | 0.012  | -0.197 | 0.053  | 0.129  | 0.115  | -0.159 | 0.136  | -0.039 | -0.014 |
| DrP |       |       | 1.000 | 0.361 | 0.093 | -0.007 | -0.375 | 0.118  | 0.321  | 0.289  | -0.335 | 0.295  | -0.124 | 0.032  |
| NoP |       |       |       | 1.000 | 0.141 | -0.006 | -0.360 | 0.122  | 0.325  | 0.306  | -0.323 | 0.325  | -0.134 | 0.057  |
| TBS |       |       |       |       | 1.000 | -0.044 | -0.148 | 0.085  | 0.151  | 0.092  | -0.148 | 0.183  | -0.075 | -0.017 |
| TMR |       |       |       |       |       | 1.000  | 0.024  | 0.012  | -0.044 | 0.005  | 0.014  | 0.031  | -0.013 | 0.040  |
| Ep  |       |       |       |       |       |        | 1.000  | -0.113 | -0.345 | -0.340 | 0.356  | -0.309 | 0.115  | 0.012  |
| DeF |       |       |       |       |       |        |        | 1.000  | 0.137  | 0.093  | -0.094 | 0.152  | 0.007  | 0.022  |
| DeFR|       |       |       |       |       |        |        |        | 1.000  | 0.271  | -0.247 | 0.246  | -0.107 | 0.056  |
| DeC |       |       |       |       |       |        |        |        |        | 1.000  | -0.282 | 0.267  | -0.066 | -0.009 |
| MeN |       |       |       |       |       |        |        |        |        |        | 1.000  | -0.337 | 0.135  | 0.010  |
| PrR |       |       |       |       |       |        |        |        |        |        |        | 1.000  | -0.164 | -0.027 |
| TSR |       |       |       |       |       |        |        |        |        |        |        |        | 1.000  | 0.031  |
| RMSE|       |       |       |       |       |        |        |        |        |        |        |        |        | 1.000  |

*Figure 4: FlexEncoder parameters correlation matrix for over 1000 evaluation samples*

- Since the top RMSE parameter sets are all item-based with *PI* value is [1, 0], we want to check the model performance on user-based ratings with *PI* value of [0, 1]. Unlike item-based rating, user-based rating shows consistent parameter patterns. The most common parameters that help achieve the RMSE in the range of 0.90 to 0.91 are: ADAM optimizer, SELU activation function and 1 round of *dense-refeeding* with *mean normalization* enabled.

## 4.7 Influence Transferability

After more than 1,000 experiments with different combinations of parameters, we want to know if the top parameter sets can generalize its prediction accuracy to bigger datasets in the same family which are MovieLens 1M and 20M. At current state, we did a preliminary testing on the MovieLens 1M dataset. The results are shown in Figure 5. We used the top 10 parameter sets in Table 2 and evaluated each of them on the MovieLens 1M. In general, they do not provide as good RMSE loss values as the 100K dataset. However, all RMSE loss values are lower than 0.98 with mean of 0.961 and the best RMSE parameter set for 100K dataset (set 1) also provides the best RMSE loss for 1M dataset with RMSE loss at 0.894. These results are remarkable since we just applied the same model architecture of the dataset 100K to dataset 1M without going through many experiments to find a good set of parameters. From this observation, we conclude that FlexEncoder model does generalize well to bigger datasets to a certain extent. This can help tremendously when finding a good parameter set for a much bigger dataset such as MovieLens 20M.

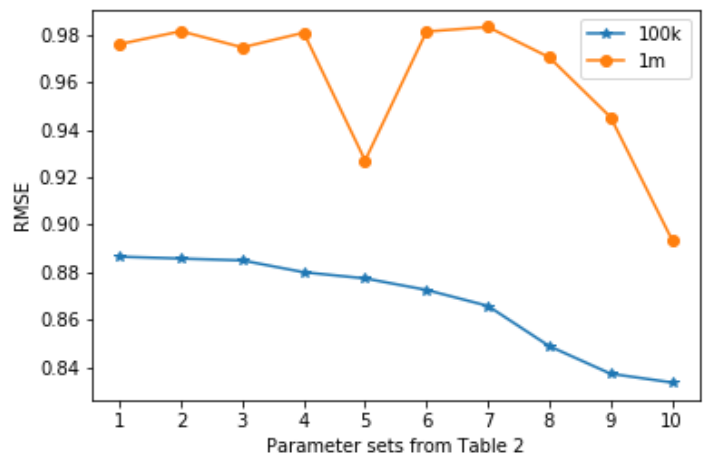

*Figure 5: RMSE results from top 10 parameter sets (Table 2) for MovieLens 100k and 1M datasets*

## 5 Conclusion

Deep Autoencoder for recommender systems shows its tremendous potential. In this work, we have presented and implemented our FlexEncoder model by embedding our own unique features and innovative features from other recent works. We argue that parameters for DAE significantly affect the model prediction accuracy and we validated this through an extensive evaluation and analysis. In addition to that, we also discover certain parameter patterns that help us to quickly identify how to tune the model to make the best out of it and to be able to achieve better accuracy than other state-of-the-art techniques. This shows the potential of our model for future development.

In the future work, we plan to improve our model by adding more parameter values such as optimization algorithm ADADELTA as well as introducing more activation functions. Additionally, we plan to test our model with other datasets such as Netflix, to evaluate how well our FlexEncoder can adapt to other datasets. Finally, we want to understand why certain sets of parameters can provide an optimal RMSE loss, which will help us in further improving our model.